\documentclass[10pt,amsmath,amssymb,aps,etoolbox,floatfix,longbibliography,nofootinbib,onecolumn,preprintnumbers,reprint,showkeys,showpacs
]
{revtex4}

\usepackage{bm}       
\usepackage{color}
\usepackage{dcolumn}  
\usepackage{enumitem}   
\usepackage{epsfig}
\usepackage{float}
\usepackage{graphicx} 
\usepackage{hyperref}
\usepackage{tensor}

\newcommand{\beq}{\begin{equation}}
\newcommand{\eeq}{\end{equation}}
\newcommand{\ds}{\displaystyle}

\begin{document}

\title{\Large 
Comment on "Acceleration of particles to high energy via gravitational repulsion in the Schwarzschild field" [Astropart. Phys. 86 (2017) 18-20]
}

\author{Alessandro D.A.M. Spallicci\textsuperscript{a,b}\footnote{
URL: http://wwwperso.lpc2e.cnrs.fr/~spallicci/}}

\affiliation{
\vskip5pt
\mbox{\textsuperscript{a}Universit\'e d'Orl\'eans}\\
\mbox{Observatoire des Sciences de l'Univers en r\'egion Centre (OSUC) UMS 3116} \\
\mbox{1A rue de la F\'{e}rollerie, 45071 Orl\'{e}ans, France}\\
\mbox{Collegium Sciences et Techniques (CoST), P\^ole de Physique}\\
\mbox{Rue de Chartres, 45100  Orl\'{e}ans, France} 
\vskip5pt
\mbox{\textsuperscript{b}Centre Nationale de la Recherche Scientifique (CNRS)}\\
\mbox{Laboratoire de Physique et Chimie de l'Environnement et de l'Espace (LPC2E) UMR 7328}\\
\mbox {Campus CNRS, 3A Avenue de la Recherche Scientifique, 45071 Orl\'eans, France}
}

\date{2 August 2017}
\begin{abstract}
Comments are due on a recent paper by McGruder III (2017) in which the author deals with the concept of gravitational repulsion in the context of the Schwarzschild-Droste solution. Repulsion (deceleration) for ingoing particles into a black hole is a concept proposed several times starting from Droste himself in 1916. It is a coordinate effect appearing to an observer at a remote distance from the black hole and when coordinate time is employed. Repulsion has no bearing and relation to the local physics of the black hole, and moreover it cannot be held responsible for accelerating outgoing particles. Thereby, the energy boost of cosmic rays cannot be produced by repulsion.  
\end{abstract}


\maketitle
\vspace{-.65 cm}
\hspace{1.55 cm}{\footnotesize Mathematics Subject Classification 2010: 83.01} 

\vspace{1.00 cm}

\section{Introduction}

In general relativity, coordinate dependence is current business. A spaceship is never seen by a distant observer to enter a black hole, while it does enter for the astronauts inside. It is a known application of the concept of repulsion - meant as deceleration - though 
it is rarely named as such. 

The first author who formulated the concept of repulsion was Droste \cite{droste1916}, who wrote the external solution of a spherical non-rotating body in May 1916 \cite{droste1916,droste1917}, independently from Schwarzschild \cite{schwarzschild1916}, both authors referring to the field equations determined by Einstein \cite{einstein1915,einstein1916a} and later through the action principle by Hilbert \cite{hilbert1917}. Incidentally, Droste already looked for a solution in 1914 \cite{droste1915} referring to the work of Einstein and Grossmann \cite{einsteingrossmann1913,einsteingrossmann1914}. To the Dutch scientist, we also owe a thorough analysis of geodesic orbits around a spherically symmetric mass and the introduction of the tortoise coordinate. There has been also a debate on the comparison between the solutions by Droste and Schwarzschild. Few tributes to Droste have appeared in the literature \cite{eisenstaedt1987, rothman2002}, but yet his name is not popular. 
We close these historical remarks taking note that repulsion emerging from the unrenormalised acceleration, i) see below, was presented in Droste's thesis  \cite{droste1916} while the repulsion from the semi-renormalised acceleration, ii) see below, in \cite{droste1917}. 

Since those early years, a discussion took place on the different geodesics that can be written using local or non-local coordinates. Large accounts  were published on this debate \cite{eisenstaedt1987,spallicci2011,spallicciritter2014,spalliccivanputten2016}. 

Two sort of claims are recurring. Ignoring previous literature, there is a long list of scholars who pretend having discovered repulsion. 
They are listed in \cite{eisenstaedt1987,spallicci2011,spallicciritter2014,spalliccivanputten2016}. 
The other concerns the attribution of physical consequences in the {\it local} environment of the falling body. 
For dismantling the latter claim, we follow the analysis by Cavalleri and Spinelli \cite{casp73, spinelli1989}.
  
Four types of measurements can be envisaged: local measurement of time $dT$, non-local measurement of time $dt$, local measurement of length $dR$, non-local measurement of length $dr$. Locality is somewhat a loose definition, but it hints at those measurements by rules and clocks affected by gravity of the Schwarzschild-Droste (SD) black hole and noted by capital letters $T,R$, while non-locality hints at measurements by rules and clocks not affected by gravity (of the SD black hole) and noted by small letters $t,r$. This definition is 
not faultless (there is no shield to gravity), but it is the most suitable to describe the debate.

For determining (velocities and) accelerations, four possible combinations do exist for ingoing particles into the black hole. 
Repulsion does not mean that the infalling particle changes direction, but that the acceleration turns positive resulting in a sort of breaking effect in the eyes of a distant observer.
  
\vskip6pt
\begin{enumerate}[label=(\roman*)]
\item{Unrenormalised acceleration $d^2r/dt^2$, distant observer adopted in \cite{mcgruder2017}; 
repulsion occurs for $ {\ds \frac {dr}{dt}} > {\ds \frac{c}{\sqrt{3}}} \left(1 - {\ds \frac{r_g}r}\right)$, or else $r <3 r_g$. }
\item{Semi-renormalised acceleration $d^2R/dt^2$, mixed coordinates observer adopted in \cite{bichma2017,chiconemashhoon2004}; 
repulsion occurs for $ {\ds \frac {dR}{dt}} > {\ds \frac{c}{\sqrt{2}}} \sqrt{1 - {\ds \frac{r_g}r}}$, or else $r <2 r_g$. }
\item{Renormalised acceleration $d^2R/dT^2$ (local observer);} 
\item{Semi-renormalised acceleration $d^2r/dT^2$.}
\end{enumerate}

The former two definitions present repulsion at different conditions of velocity and acceleration. They share the common feature of referring to coordinate time. The third definition recurs to proper time and proper length and it is never repulsive. First and third make use of coordinates of an unique nature, either local or remote. The fourth definition has been disregarded in the literature concerning repulsion\footnote{This might be due to the difficulty of achieving a reasonable transmission of length, conversely to the delays taken by the signals to reach the distant clock which do not change in a static field.}.


The references \cite{eisenstaedt1987,spallicci2011,spallicciritter2014,spalliccivanputten2016} show the conditions on velocity or accelerations for which repulsion takes place. Furthermore, references \cite{spallicciritter2014,riaospco2016b} consider also the effect of self-force on radial fall. 

\section{Comments on the paper}

Having recalled the science scenario in the previous section, a list of comments on the paper can be now issued. 

\begin{enumerate}
{\item Equation (3) in \cite{mcgruder2017} appeared first in \cite{droste1916} and it is present in classical textbooks. It corresponds to the unrenormalised acceleration $d^2r/dt^2$.}
{\item Similar figures to Figs. 1-3 in \cite{mcgruder2017}, have appeared in \cite{casp73, spinelli1989}.} 
{\item The author of \cite{mcgruder2017} appears to apply repulsion to outgoing particles. The latter might be subjected also to a maximal velocity and acceleration, as the incoming particles, but talking of repulsion is epistemologically inappropriate. Repulsion by what? The particles leave the black hole and continue doing so if their initial energy allows it.}
{\item The author of \cite{mcgruder2017} considers repulsion as a sort of energy booster for outgoing particles, namely cosmic rays, without justification in our opinion. Indeed, Fig. 3 in \cite{mcgruder2017} has to be read from infinity to the black hole. We see that at the horizon, the infalling particle slows down and never enters the horizon. But in the opposite sense,  the black  hole does not push away and speed up the outgoing particle.} 
\item{The main shortcoming of \cite{mcgruder2017} is the assumption that repulsion, type i), is a physical effect and not just a coordinate effect.  
In other words, repulsion is only present when coordinate time is adopted and cannot explain phenomena local to the source, and ruled by proper time\footnote{The same remark can be opposed to findings in \cite{bichma2017} and references therein. Interestingly, the difference lies in adopting type another type of repulsion, namely ii) \cite{chiconemashhoon2004}. }.}\\
\vskip 6pt
Minor remarks:
\item{The author does not use $c=1$, see Eq. (2); thereby, in Eq. (1), dt should be read as $c^2dt^2$; in Eq. (3) $dr/dt$ should be read as $dr/cdt$; Eq. (7) should be multiplied by $c$.}
\item{Eq. (6) is derived for $r\rightarrow \infty$ from Eq. (5) and not Eq. (3).}
\item{There are editorial inaccuracies in the references. }
\end{enumerate}

\bibliography{references_spallicci_170802}

\end{document}